# Energy gaps of atomically precise armchair graphene nanoribbons

Wen-Xiao Wang[1], Mei Zhou[2], Xinqi Li[3], Si-Yu Li[1], Xiaosong Wu[3], Wenhui Duan[2], and Lin He[1,*]

[1]Center for Advanced Quantum Studies, Department of Physics, Beijing Normal University, Beijing, 100875, People's Republic of China

[2]State Key Laboratory of Low-Dimensional Quantum Physics and Collaborative Innovation Center of Quantum Matter, Department of Physics, Tsinghua University, Beijing, 100084, People's Republic of China

[3]State Key Laboratory for Artificial Microsctructure and Mesoscopic Physics, Peking University, Beijing 100871, China and Collaborative Innovation Center of Quantum Matter, Beijing 100871, China

* Email: helin@bnu.edu.cn

**Graphene nanoribbons (GNRs) are one-dimensional (1D) structures that exhibit a rich variety of electronic properties[1-17]. Therefore, they are predicted to be the building blocks in next-generation nanoelectronic devices. Theoretically, it has been demonstrated that armchair GNRs can be divided into three families, i.e., $N_a = 3p$, $N_a = 3p + 1$, and $N_a = 3p + 2$ (here $N_a$ is the number of dimer lines across the ribbon width and $p$ is an integer), according to their electronic structures, and the energy gaps for the three families are quite different even with the same $p$[1,3-6]. However, a systematic experimental verification of this fundamental prediction is still lacking, owing to very limited atomic-level control of the width of the armchair GNRs investigated[7,9,10,13,17]. Here, we studied electronic structures of the armchair GNRs with atomically well-defined widths ranging from $N_a = 6$ to $N_a = 26$ by using scanning tunnelling microscope (STM). Our result demonstrated explicitly that all the studied armchair GNRs exhibit semiconducting gaps due to quantum confinement and, more importantly, the observed gaps as a function of $N_a$ are well grouped into the three categories, as predicted by density-functional theory calculations[3]. Such a result indicated that we can tune the electronic properties of the armchair GNRs dramatically by simply adding or cutting one carbon dimer line along the ribbon width.**

The electronic structures of GNRs are strongly dependent on their widths and edge orientation[1-6]. For example, the energy gaps of the armchair GNRs sensitively depend on the number of dimer lines $N_a$ along the ribbon width, as shown in Fig. 1 (see Methods for details of first principles calculations). For the three different categories with $p = 5$, the armchair GNRs with $N_a = 15$ ($N_a = 3p$) and $N_a = 16$ ($N_a = 3p + 1$) display energy gaps $E_g = 556$ meV and $E_g = 657$ meV respectively, whereas, the armchair GNR with $N_a = 17$ ($N_a = 3p + 2$) is predicted to exhibit a much smaller gap $E_g = 118$ meV (Fig. 1b). Because of the highly tunable feature of their electronic properties, armchair GNRs are believed to be one of the most promising candidates toward the design of graphene-based circuits for technological applications. This stimulates the development of many different methods in fabricating high-quality GNRs[10,12,13,17-19]. However, the studied armchair GNRs with atomically well-defined widths and edge orientation are rather limited up to now. A rare example is the observation of about 100 meV bandgap in a $N_a = 5$ armchair GNR (the length of the studied GNR is about 5 nm), which really indicates that the $N_a = 3p + 2$ (here $p = 1$) armchair GNR should exhibit a very small bandgap[17]. Very recently, STM study of armchair GNRs with widths ranging from about (3.5 ± 0.5) nm to (10.5 ± 0.5) nm revealed that these GNRs can be grouped into two families: one displays large gaps, which is attributed to the ribbons belonging to the $N_a = 3p$ and $N_a = 3p + 1$ classes; the other exhibits no detectable gap, which is attributed to the ribbons belonging to the $N_a = 3p + 2$ class[13]. This result is clearly indicative of width-dependence physics in the armchair GNRs, however, the lack of exact number of carbon dimer lines $N_a$ across the width of these GNRs, owing to the measurement error and the atomic-scale edge irregularities[13], did not allow a more systematic insight.

To experimentally verify the width-dependence physics in the armchair GNRs, we realized the growth of 0.6-3.1 nm wide armchair GNRs with well-defined $N_a$ ($6 \leq N_a \leq 26$) on the nanometer-wide C-face terraces of SiC (see Methods for details)[20-22], as schematically shown in Fig. 2a. These ultra-narrow GNRs provide us unprecedented opportunities to address this issue and, moreover, our scanning probe technique allows us to directly measure the energy gaps and the exact $N_a$ of these GNRs simultaneously. Previously, the growth of GNRs on the slope of a Si-face terraced SiC surface with crystallographically perfect edges was reported[14-16]. These GNRs are usually of several tens nanometers and the boundaries of the GNRs are defined by regions where they attach to graphene tightly bonded to the (0001) oriented terraces (in these terraces, the carbon atoms are strongly bonded to the exposed Si atoms of SiC), i.e., the so-called graphene buffer layer of Si-face of SiC[14,15]. It has been demonstrated that the GNRs on the slope are decoupled from the substrate and behave as free-standing ribbons with similar widths[14-16]. Very recently, 1-2 nm wide GNRs decoupled from the substrate were further observed on the sidewall between the (0001) oriented nanoterraces of SiC[16]. Our experimental result, as shown in Fig. 2, demonstrates that similar ultra-narrow GNRs could also be formed on the C-face nanoterraces of SiC.

The scanning tunnelling spectroscopy (STS) measurements in Fig. 2c indicate that graphene sheet in the C-face terraces (Fig. 2b) is decoupled from the substrate, as reported previously (see Methods for details of STM measurements). The spectra of the graphene sheet, recorded in the magnetic field of 4 T, exhibit Landau quantization of massless Dirac fermions (with the hallmark zero-energy Landau level and its characteristic nonequally spaced energy-level spectrum of Landau levels), as expected to be observed in a pristine graphene monolayer[23-26]. In between the C-face nanoterraces, the graphene is strongly bonded to the substrate, i.e., the exposed Si atoms of SiC, and behaves as the graphene buffer layer. The profile line across the C-face nanoterraces (Fig. 2f) directly demonstrated that the freestanding graphene regions on the nanoterraces and the pinning regions on the sidewall are quite different, owing to their different coupling strengths with the substrate. The spatial variation of the interaction with the substrate can change the electronic properties of graphene dramatically[14-16,27]. The ultra-narrow armchair GNRs on the C-face nanoterraces, as shown in Fig. 2d and 2e, are expected to exhibit energy gaps due to quantum confinement. Theoretically, the energy gaps of these decoupling armchair GNRs on SiC surface, obtained by first principle calculations, are on the order of magnitude of the expected gaps for free-standing armchair GNRs with similar widths[16]. In our experiment, the widths of the armchair GNRs, i.e., the number of carbon dimer lines across the ribbons width, can be determined exactly according to the atomic-resolution STM images (Fig. 2e) and the profile line across the ribbons (Fig. 2f). After carefully examining hundreds of C-face nanoterraces on SiC surface, we obtain armchair GNRs with different widths and the measured value of $N_a$ ranges from $N_a = 6$ to $N_a = 26$. Therefore, these armchair GNRs provide an attractive platform to explore the highly tunable semiconductor bandgaps, which are simply determined by the value of $N_a$, in the armchair GNRs.

Figure 3 shows several representative STS spectra of the armchair GNRs with different $N_a$. Obviously, all the studied armchair GNRs exhibit semiconducting gaps. For clarity, we plotted the spectra according to the three categories of the widths of the armchair GNRs. For each category, the energy gaps of the GNRs decrease with increasing $N_a$, as expected to be observed due to the quantum confinement. A notable feature of the spectra is that the $N_a = 3p + 2$ armchair GNRs display a much smaller gap comparing to that of the other two categories. Such a result demonstrated directly that the energy bandgaps of the armchair GNRs depend sensitively on the value of $N_a$.

Figure 4 summarized the energy bandgaps $E_g$ of the studied armchair GNRs as a function of $N_a$. The energy gaps of freestanding armchair GNRs with different widths obtained by first principles theoretical calculations[3] are also plotted for comparison. Obviously, the theoretical result reproduces the overall features of the observed width dependent energy gaps for the armchair GNRs on SiC surface. Such a result agrees with previous theoretical calculations that the armchair GNRs on SiC surface behave as free-standing armchair GNRs[16]. The observed gaps as a function of $N_a$ can be well grouped into the three categories predicted by first principles calculations. This result demonstrated explicitly that the energy gaps of these decoupling armchair GNRs on SiC surface are on the order of magnitude of the expected gaps for free-standing armchair GNRs with similar widths. The $E_g \sim N_a^{-1}$ behavior for all the three categories further confirms that the origin of the energy gaps for GNRs with armchair edges is the quantum confinement.

The above result indicates that we can tune the electronic structures of the armchair GNRs dramatically by simply adding or cutting one dimer line along the ribbon width. For an armchair GNR with a fixed $N_a$, it exhibits an almost constant energy bandgap according to the spatially resolved tunnelling spectra along the ribbon (see Fig. 5a for an example). In Fig. 4, we only summarized the result of the armchair GNRs with a fixed $N_a$. However, in some of the armchair GNRs on SiC surface, spatially resolved tunnelling spectra, as shown in Fig. 5b and Fig. 5c for examples, reveal that the energy gaps vary spatially along the ribbon. Such a result is attributed to the width-dependent energy gaps of the armchair GNRs. For the nanometer-wide armchair ribbons, the energy gaps vary a lot even when we only change one carbon dimer line across the ribbon width, as shown in Fig. 5c. This behavior not only provides us unprecedented control over electronic properties of the graphene GNRs at atomic-level, but also opens the way towards the realization of electronic junctions entirely in a graphene GNR.

**Methods:**

**(a.1) STM and STS measurements.** The scanning tunneling microscopy (STM) system is an ultrahigh vacuum single-probe scanning probe microscope (USM-1500) from UNISOKU. All STM and scanning tunneling spectroscopy (STS) measurements were performed at liquid-helium temperature and the images were taken in a constant-current scanning mode. The STM tips were obtained by chemical etching from a wire of Pt(80%) Ir(20%) alloys. Lateral dimensions observed in the STM images were calibrated using a standard graphene lattice as well as a Si (111)-(7×7) lattice and the STS spectra were calibrated using a Ag (111) surface. The STS spectrum, i.e., the dI/dV-V curve, was measured with a standard lock-in technique by applying alternating current modulation of the bias voltage of 7 mV (871 Hz) to the tunneling bias.

**(a.2) Samples Growth on SiC**: We prepared a high-quality graphene film on a silicon carbide crystal by thermal decomposition process. On-axis semi-insulating 4H-SiC wafers were purchased from Cree Inc, and the mis-cut angle was estimated to be 5′. Prior to growth, SiC chips were hydrogen etched at 1600 °C for 20 min to remove polishing scratches, so we obtained atomically flat SiC substrate surface. The growth was carried out in a home-made high vacuum induction furnace. Samples were first annealed in vacuum at about 1000 °C for 60 min to remove the native oxide on the surface. Then growth took place in a mixed gas flow (2% hydrogen and 98% argon) at 20-40 sccm. The hydrogen partial pressure is about 0.03 mbar. After 15 min of growth at about 1500 °C, heating was shut off and the samples were allowed to cool naturally.

**(a.3) Computational Methods:** The first-principles calculations of electronic properties were carried out within the framework of the density functional theory (DFT) using local density approximation (LDA)[28] exchange-correlation functional with projector augmented-wave (PAW)[29,30] method as implemented in Vienna Ab-initio Simulation Package (VASP)[31]. The energy cutoff for the plane-wave basis is 400 eV. The structures of the studied $N_a$ armchair GNR are fully relaxed until the residual force is less than 0.01 eV/Å per atom. A vacuum region of about 20 Å was built to avoid the interaction between planes and edges. A 1× 32× 1 Γ-centered k-point mesh was adopted in the self-consistent calculations.

**Acknowledgements**

This work was supported by the National Basic Research Program of China (Grants Nos. 2014CB920903, 2013CBA01603), the National Natural Science Foundation of China (Grant Nos. 11422430, 11374035, 11334006, 11222436), the program for New Century Excellent Talents in University of the Ministry of Education of China (Grant No. NCET-13-0054), Beijing Higher Education Young Elite Teacher Project (Grant No. YETP0238). L.H. also acknowledges support from the National Program for Support of Top-notch Young Professionals. The first principles calculations were performed on the "Explorer 100" cluster system at Tsinghua University.

**Author contributions**

L.H. conceived and provided advice on the experiment and analysis. W.D. conceived and provided advice on DFT calculation. W.X.W. performed the experiments and analyzed the data. M.Z. performed the theoretical calculations. T.C. and X.W. synthesized the graphene on SiC substrate. L.H. wrote the paper. All authors participated in the data discussion.

**Competing financial interests:** The authors declare no competing financial interests.

**Figure Legends**

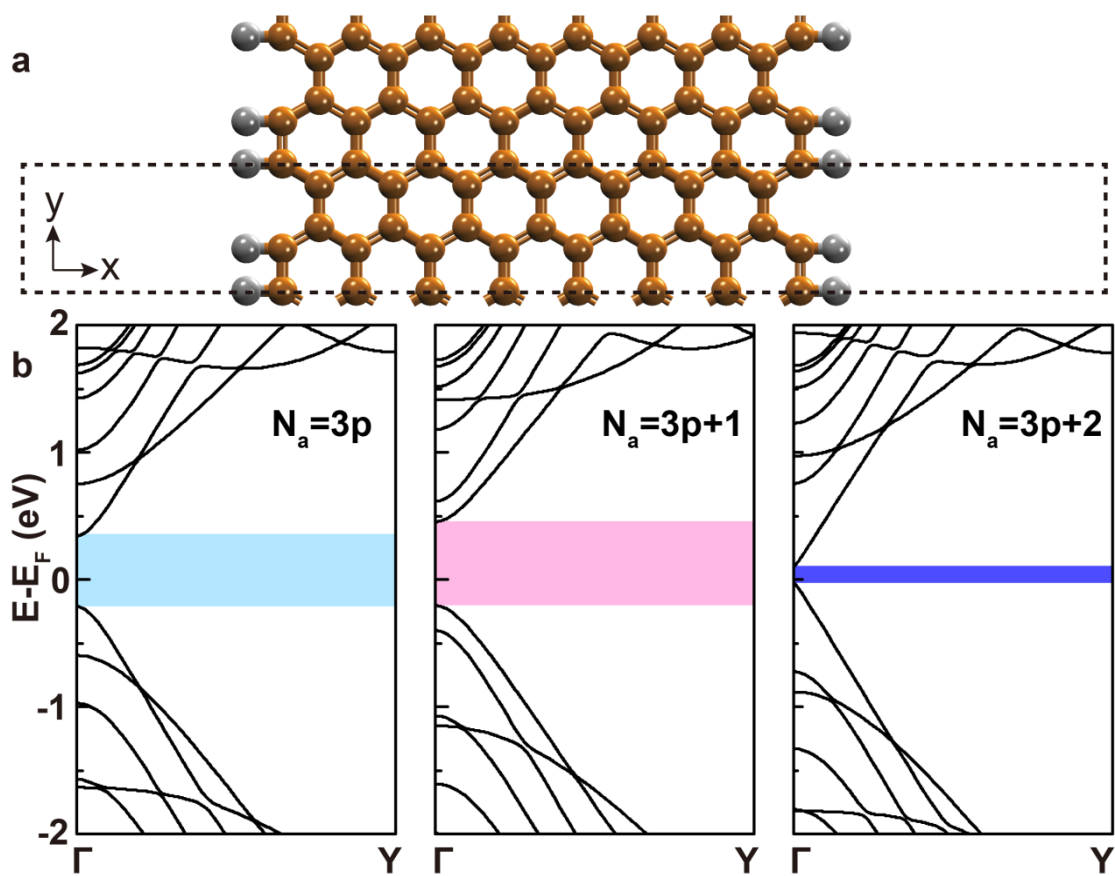

**Figure 1. Geometry and electronic structure of armchair GNRs. a**. Schematic of a $N_a = 15$ armchair GNR with the edge dangling σ bonds passivated by hydrogen atoms (Grey balls). Black dashed frame indicates the supercell in the calculation. **b**. Band structures of $N_a$ armchair GNRs with $N_a = 3p$, $3p + 1$ and $3p + 2$ (here $p = 5$) obtained by first principles calculations. The coloured areas indicate the energy bandgaps of the three different families of the armchair GNRs. Obviously, the $N_a = 3p + 2$ armchair GNR exhibits a much smaller bandgap.

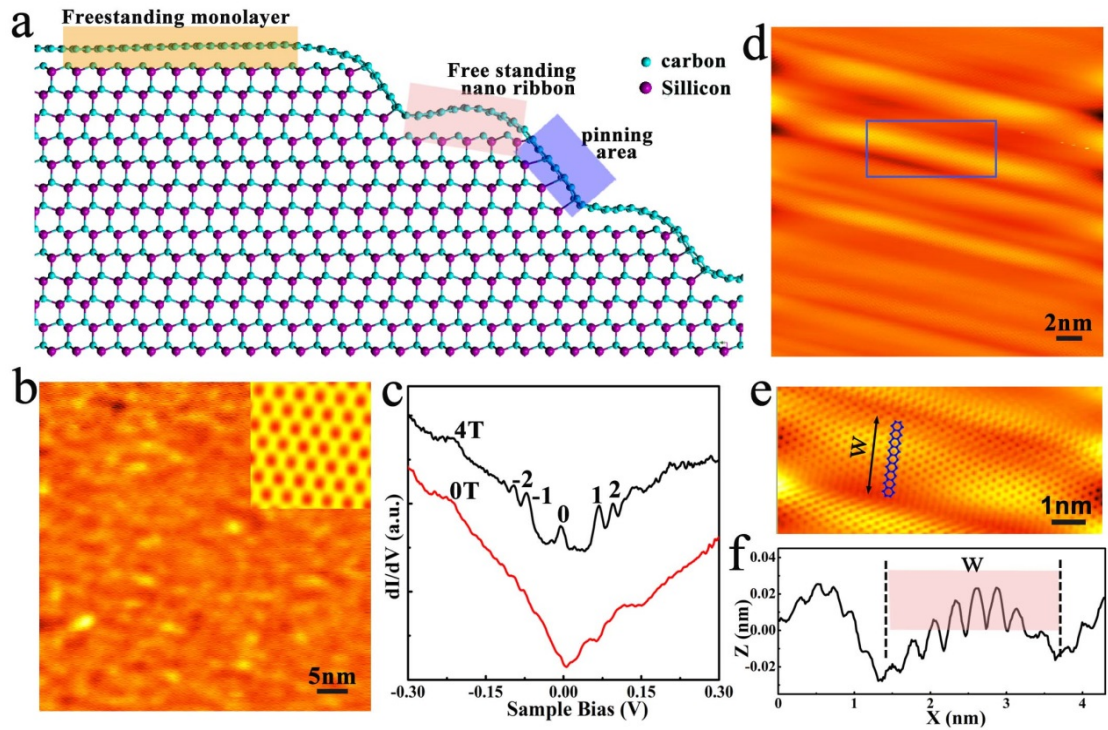

**Figure 2. Structures of graphene and GNRs on C-face terraces of SiC surface. a**. Schematic structure of graphene monolayer on C-face terraces of SiC surface. There are three different regions exhibiting quite different electronic properties. **b**. A representative STM image of a graphene sheet on a C-face terrace of SiC surface ($V_{sample}$ = - 0.9 V, $I$ = 0.28 nA). The inset shows an atomic-resolution STM image of the graphene sheet. **c**. STS spectra, i.e., *dI/dV-V* curves, taken at the position in panel (b) under zero and 4 T magnetic fields. For clarity, the curves are offset on the Y-axis and Landau level indices of massless Dirac fermions are marked. **d**. A typical STM image ($V_{sample}$ = 0.8 V, $I$ = 0.3 nA) showing a graphene sheet across several adjacent C-face nanoterraces of SiC surface. The atomic-resolution STM image of an armchair GNR on a C-face nanoterrace is shown in panel **e**. The atomic structure of graphene is overlaid onto the STM image to determine the number of carbon dimer lines along the ribbon width. **f**. A profile line across the GNR width.

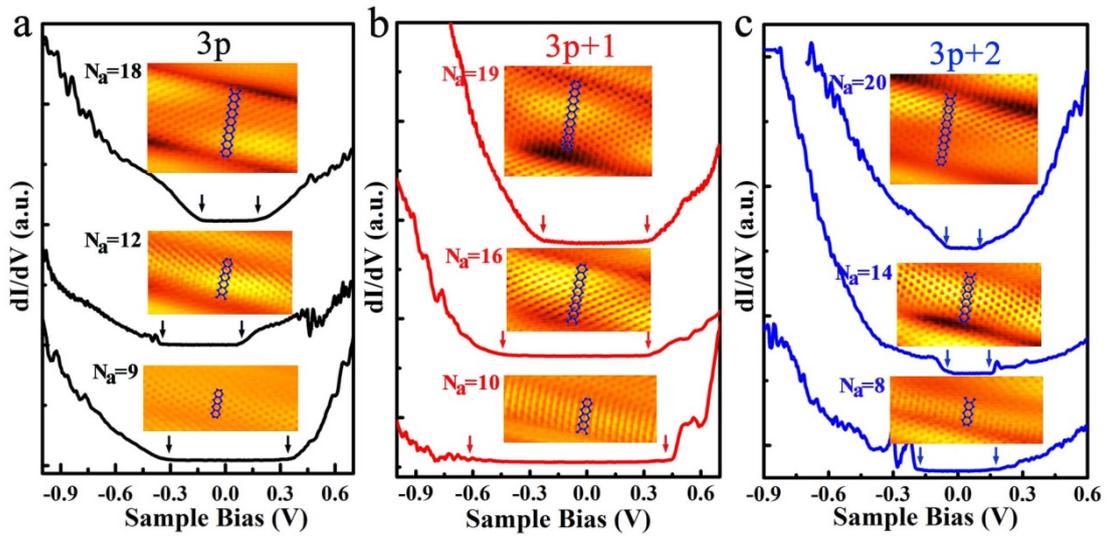

**Figure 3. STS spectra of armchair GNRs with different widths on C-face nanoterraces of SiC surface.** Representative spectra of different armchair GNRs belonging to the (**a**) $N_a = 3p$, (**b**) $N_a = 3p + 1$, and (**c**) $N_a = 3p + 2$ families. The corresponding atomic-resolution STM images of these armchair GNRs are shown in the insets. The edges of the energy bandgap of the STS spectra are marked by two arrows.

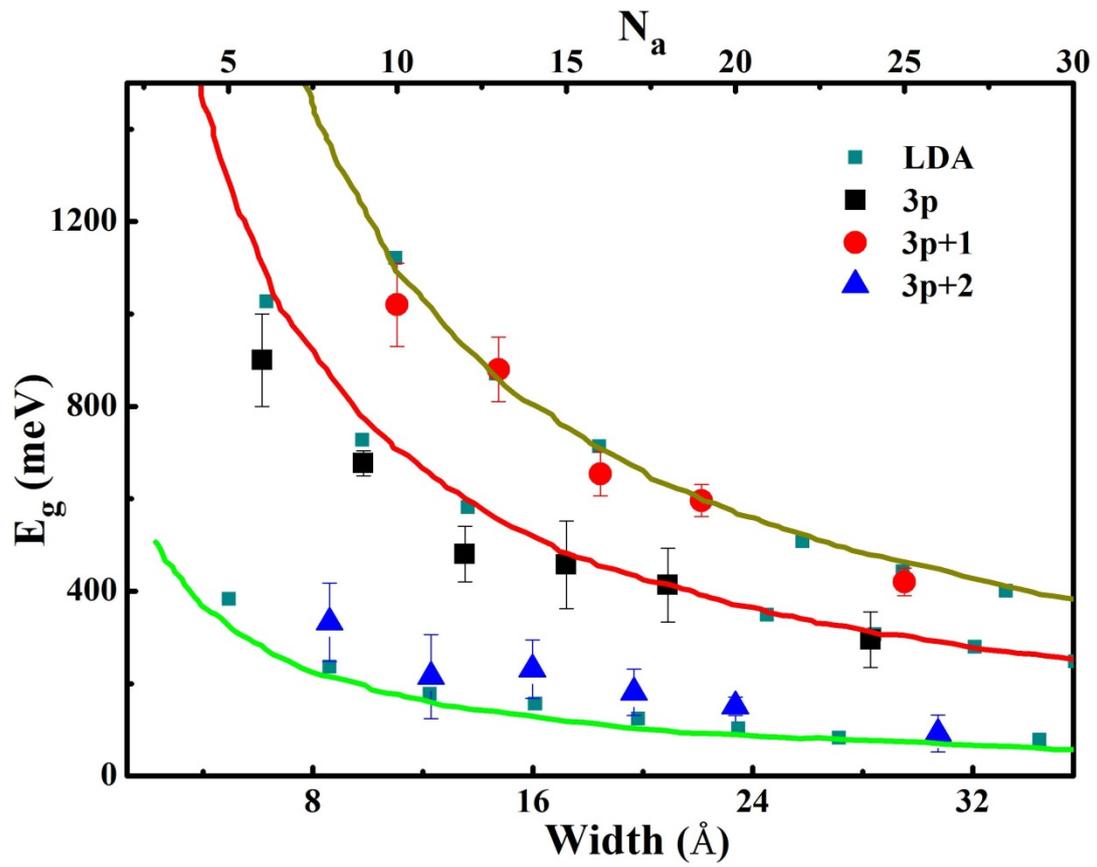

**Figure 4. The measured bandgap of the armchair GNRs as a function of widths.** The measured energy bandgaps can be grouped into three categories. The theoretical data points taken from ref. 3 are also given for comparison.

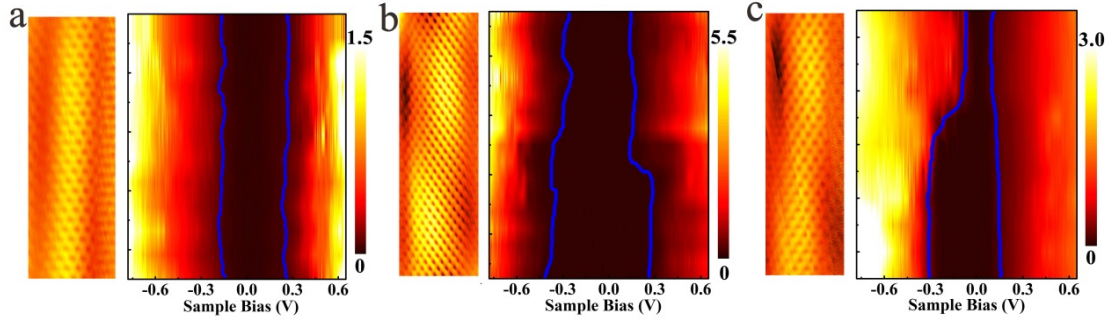

**Figure 5. Spatially resolved band gaps of three armchair GNRs by tunnelling spectroscopy. a.** Left: Atomic resolution STM image of an armchair GNR with $N_a = 3p + 2 = 8$. Right: Intensity plot of d$I$/d$V$ spectra taken along the armchair GNR. The armchair GNR exhibits a constant bandgap along it. **b.** Left: Atomic resolution STM image of an armchair GNR showing a transition of widths from $N_a = 3p = 15$ (top) to $N_a = 3p + 1 = 16$ (bottom). Right: Intensity plot of d$I$/d$V$ spectra taken along the armchair GNR exhibiting a slight variation of the gaps along it. **c.** Left: Atomic resolution STM image of an armchair GNR showing a transition of widths from $N_a = 3p + 2 = 11$ (top) to $N_a = 3p = 12$ (bottom). Right: Intensity plot of d$I$/d$V$ spectra taken along the armchair GNR exhibiting a large variation of the gaps along it. The edges of bandgap are highlighted with blue lines in the spectra maps.